\documentclass[english,prl,aps,reprint,doublecolumn,superscriptaddress,longbibliography]{revtex4-2}
\usepackage[T1]{fontenc}
\usepackage[latin9]{inputenc}
\usepackage{bm}
\usepackage{float}
\usepackage{amsmath}
\usepackage{amssymb}
\usepackage{graphicx}

\makeatletter

\providecommand{\tabularnewline}{\\}

\makeatother

\usepackage{babel}
\begin{document}
\title{A Solvable Model for Discrete Time Crystal Enforced by Nonsymmorphic
Dynamical Symmetry}
\author{Zi-Ang Hu}
\affiliation{Department of Physics, The University of Hong Kong, Pokfulam Road,
Hong Kong, China}
\author{Bo Fu}
\affiliation{Department of Physics, The University of Hong Kong, Pokfulam Road,
Hong Kong, China}
\author{Xiao Li}
\affiliation{Department of Physics, City University of Hong Kong, Kowloon, Hong
Kong, China}
\affiliation{City University of Hong Kong Shenzhen Research Institute, Shenzhen
518057, Guangdong, China}
\author{Shun-Qing Shen}
\email{sshen@hku.hk}

\affiliation{Department of Physics, The University of Hong Kong, Pokfulam Road,
Hong Kong, China}
\date{\today}
\begin{abstract}
Discrete time crystal is a class of nonequilibrium quantum systems
exhibiting subharmonic responses to external periodic driving. Here
we propose a class of discrete time crystals enforced by nonsymmorphic
dynamical symmetry. We start with a system with nonsymmorphic dynamical
symmetry, in which the instantaneous eigenstates become M\"{o}bius twisted,
hence doubling the period of the instantaneous state. The exact solution
of the time-dependent Schr\"{o}dinger equation shows that the system spontaneously
exhibits a period extension without undergoing quantum superposition
states for a series of specific evolution frequencies or in the limit
of long evolution period. 
Moreover, in such case the system gains a $\pi$ Berry
phase after two periods' evolution. 
Finally, we show that the subharmonic response is stable even when many-body interactions are introduced, indicating a DTC phase in the thermodynamic limit.  
\end{abstract}
\maketitle

\paragraph*{Introduction}

Recently, the spontaneous breaking of time translation symmetry has
attracted tremendous attention and led to the idea of time crystal~\citep{WilczekPrl12}.
Although it has been shown that no system can spontaneously break
the continuous time translation symmetry~\citep{WatanabePrl15},
it is possible to break the discrete time translation symmetry in
Floquet quantum many-body systems~\citep{ElsePrl16,ElsePrx16,Sachapra2015,SachaRpp2017,YaoPrl17,YaoNp2020,Russomannoprb17,Shapereprl12,TaherNCi2022,Suraceprb2019,liang2020time,pizzi2021higher}.
This new nonequilibrium phase has since been verified in a series
of experiments~\citep{Choinature2017,Jurcevicprl17,RovnyPrl18,zhang2017observation,NeufeldNc2019,KyprianidisScience21,wang2022observation,PhilippSciadv2022}.
Subsequent studies in nonequilibrium Floquet quantum systems further
generalize to incorporate both spatial and temporal dimensions, giving
rise to a wide range of new phenomena, including the space-time crystals~\citep{XuPrl18,Gaoprl2021,Peng2022prl,Smitsprl2018,Giergielprl21}.
In particular, it has been shown that a nonsymmorphic symmetry can
enforce energy band crossings and create a M\"{o}bius twist as well
as additional topological properties~\citep{MichelPrb99,FuNpjQ2022,ZhaoPrb16,QizhouPra18,ShiozakiPrb15,Hollerprb2018,Giergielprl21}.
However, it remains unclear what will happen when a system hosts nonsymmorphic
\textbf{\textit{dynamical}} symmetry.

Here we propose a class of discrete time crystals (DTCs) enforced
by the nonsymmorphic dynamical symmetry. We reveal the dynamical behavior
of such systems by presenting an exact solution of the time-dependent
evolution of a two-level system. The instantaneous eigenstates of
the system are M\"{o}bius twisted when the nonsymmorphic dynamical symmetry
is present. The exact time-dependent solution shows that the quantum
states of the system exhibit evolution along the twisted instantaneous
state so that period of the states is doubled to that of the system
Hamiltonian when the ratio between the width of the instantaneous
energy band to the evolution frequency of the system's Hamiltonian
equals a set of discrete values, or when the system evolves in the
long-period limit. Besides, we find that the system has a nontrivial
topology with a half-integer winding number and a $\pi$ Berry phase
in two cycles of evolution. Also, we explore the system's stability
in the presence of many-body interactions and show the robustness of this class
of DTCs. The exact solution provides an explicit evidence to support
the spontaneous breaking of the time translation symmetry in discrete
time crystals.

\paragraph*{Model with a M\"{o}bius twist in instantaneous state}

\begin{figure}[htbp]
\includegraphics[width=8.5cm]{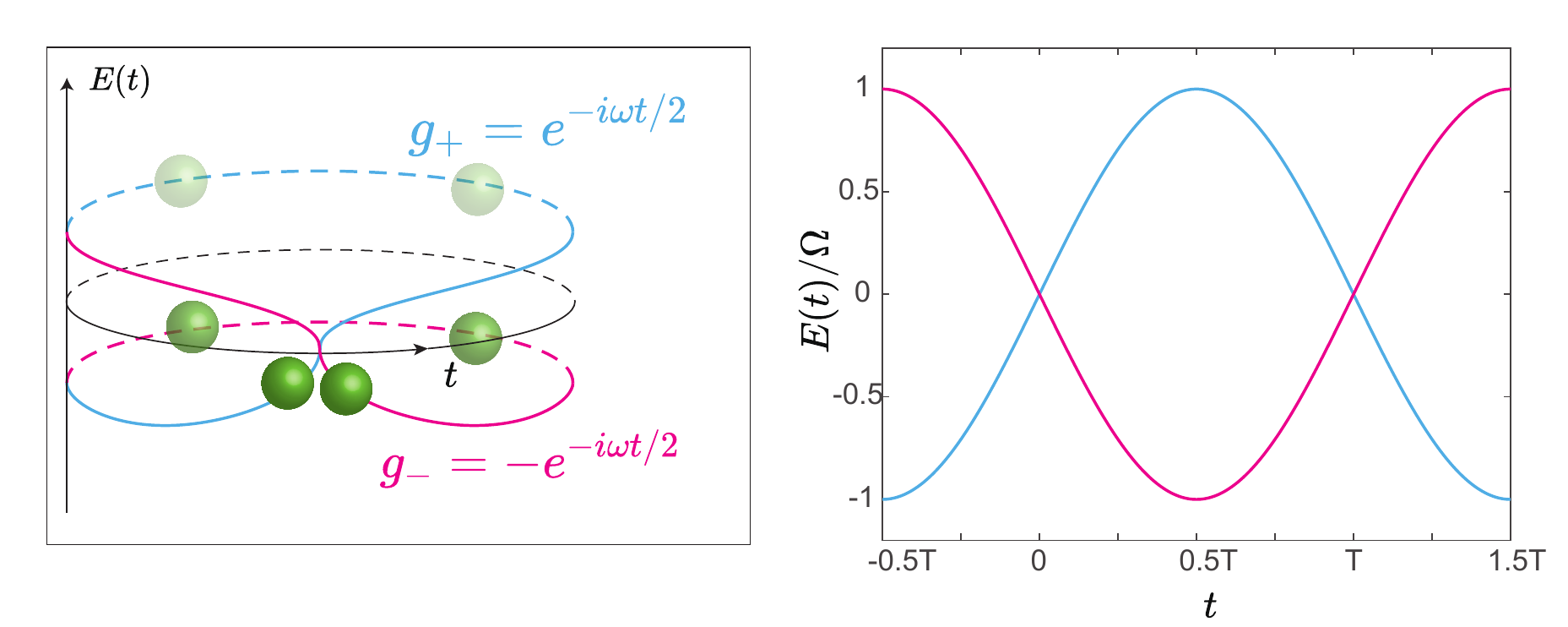} \caption{\label{MobTwist} Illustration of M\"{o}bius twist of the instantaneous
eigenstate and evolution process of DTC. 
The red and blue lines label different
eigenstates of the symmetry operator $G$. The vertical direction shows
the relativity energy of two states. The two eigenstates will swap
after one period of evolution and go back to themselves after two
periods. The green ball represent the quantum states evolution which
will go into a superposition state of instantaneous states (translucent
balls) and go to another instantaneous state after one period evolution.}
\end{figure}

The nonsymmorphic dynamical symmetry operation represents a unique
category of symmetry operations that integrates both spatial transformations
and non-trivial time translation operations, which cannot be performed
independently. When nonsymmorphic symmetry is incorporated, the resulting
extension of the symmetry group exhibits topologically non-trivial
properties, leading to intriguing characteristics in the group representations.
In the realm of group representation theory, the instantaneous eigenstate
of the system adheres to the irreducible representation of the symmetry
group, while the Hamiltonian aligns with the induced representation
of the nonsymmorphic symmetry. When the system exhibits nonsymmorphic
symmetry, the period of the irreducible representation does not consistently
align with the period of the induced representation. Consequently,
this inconsistency generates a mismatch between the periods of the
instantaneous state and the Hamiltonian.

Specifically, we start with a time-dependent two-level Hamiltonian
with the dynamical glide symmetry \citep{FuNpjQ2022}, 
\begin{equation}
H_{0}(t)=\frac{1}{2}\hbar\Omega\sin(\omega t)\sigma_{x}+\hbar\Omega\sin^{2}(\frac{\omega t}{2})\sigma_{y},\label{eq:Ham}
\end{equation}
where $\sigma_{i}$ ($i=x,y,z$) are the Pauli matrices. The Hamiltonian
has a period of $T=\frac{2\pi}{\omega}$. At time $t$, the instantaneous
eigenstates are $\phi_{\chi}(t)=\frac{1}{\sqrt{2}}[\chi e^{-i\omega t/2},1]^{\mathsf{T}}$
($\mathsf{T}$ indicating the transpose), with $\chi=\pm1$. The corresponding
energy eigenvalues are $E_{\chi}(t)=\chi\hbar\Omega\sin(\omega t/2)$.
The model possesses a nonsymmorphic dynamical symmetry $\tilde{g}$,
the square of which will be a time translation symmetry with one period.
The matrix representation of $\tilde{g}$ in the basis of the two-level
system is the reduced representation of the whole group and has the
following form 
\begin{equation}
G(t)=\begin{pmatrix}0 & e^{-i\omega t}\\
1 & 0
\end{pmatrix}.
\end{equation}
As a result, the Hamiltonian commutes with $G(t)$, $[H(t),G(t)]=0$.
Because $G(t)$ is the induced representation, it has the same period
$T$ as the Hamiltonian, $G(t+T)=G(t)$. Furthermore, $G^{2}(t)=e^{-i\omega t}$
is the irreducible representation of the time translation symmetry
with a period $T$. Consequently, the eigenvalues of $G(t)$ are $g_{\chi}(t)=\chi e^{-i\omega t/2}$.
The corresponding eigenstates are the instantaneous eigenstates of
the system $\phi_{\chi}(t)$. The eigenvalues $g_{\chi}(t)$ are in
the irreducible representations of the extended symmetry group whose
period is twice the system's period. In addition, it has the swapping
property within one period as $g_{\chi}(t+T)=g_{-\chi}(t)$ and $g_{\chi}(t+2T)=g_{\chi}(t)$.
The corresponding eigenstates $\phi_{\chi}(t)$ share the same properties.
The swapping property of the eigenvalues and eigenstates is a manifestation
of the group monodromy. It directly illustrates the periodic extension
by introducing the nonsymmorphic symmetry. Another consequence of
this period mismatch is that the instantaneous eigenstates of the
system automatically become M\"{o}bius twisted. Figure~\ref{MobTwist}
illustrates how the M\"{o}bius twist happens. The two colored lines represent
the two eigenstates of the symmetry operator $G(t)$, which are also
the instantaneous eigenstates of the Hamiltonian. After one cycle
of evolution, the two eigenstates will swap. Since the instantaneous
eigenstates of the Hamiltonian still host the discrete time translation
symmetry, it is forced to undergo a symmetry-enforced crossing in
the evolution process. The two eigenstates are M\"{o}bius twisted at $t=T$
such that they only come back to themselves after a two-period evolution,
matching the period of the irreducible representation of the symmetry
group.

Besides period doubling in the two-level system with a dynamical glide
symmetry, systems with other nonsymmorphic symmetries can also have
other forms of period extension. For example, a dynamical screw symmetry
in multi-level systems allows the period extension of more than two~\citep{sup}.

\paragraph*{Exact solution and DTC}

The exact solution $\Psi(t)$ of the time-dependent Hamiltonian in
Eq.~(\ref{eq:Ham}) can be obtained by solving the time-dependent
Schr\"{o}dinger equation, $i\hbar\partial_{t}\Psi(t)=H_{0}(t)\Psi(t)$.
In the basis of the instantaneous eigenstates $\phi_{\chi}(t)$, an
ansatz for the time-dependent solution is $\Psi(t)=\sum_{\chi}c_{\chi}(t)\exp\left[i\frac{\omega t}{4}-\frac{i}{\hbar}\int_{0}^{t}E_{\chi}(t')dt'\right]\phi_{\chi}(t),$
where $c_{\chi}(t)$ are two time-dependent coefficients. Substituting
the wave function into the time-dependent Schr\"{o}dinger equation, we
find that $c_{\chi}(t)$ satisfies 
\begin{equation}
\partial_{x}^{2}c_{\chi}-i\chi\alpha\sin(2x)\partial_{x}c_{\chi}+c_{\chi}=0,\label{eq:c-eq}
\end{equation}
where $x=\frac{\omega t}{4}$ and $\alpha=\frac{8\Omega}{\omega}.$
Note that $c_{\chi}(t)$ depend on $\chi$, $\alpha$, and $x$. We
denote the initial state as $\Psi(0)=\sum_{\chi}c_{\chi}(0)\phi_{\chi}(0)$
with $c(x)\equiv(c_{+}(x),c_{-}(x))^{\mathsf{T}}$. The solutions
for $c_{\chi}(t)$ are found as follows: 
\begin{equation}
c(x)=\begin{pmatrix}H_{c}^{+}(\alpha,x) & -i\sin(x)H_{c}^{-}(\alpha,x)\\
-i\sin(x)H_{c}^{-}(\alpha,x)^{*} & H_{c}^{+}(\alpha,x)^{*}
\end{pmatrix}c(0).\label{eq:exact-c-solution}
\end{equation}
Here we define 
\begin{equation}
H_{c}^{\chi}(\alpha,x)\equiv H_{c}(i\alpha,-\frac{\chi}{2},-\frac{1}{2},-\frac{i\alpha}{2},\frac{1}{8}+\frac{i\alpha}{4};\sin^{2}x),
\end{equation}
where $H_{c}$ is the confluent Heun function~\citep{Heun1888,olver2010nist},
which satisfies the boundary conditions $H_{c}^{\chi}(\alpha,x=0)=1$
and $\frac{d}{dx}H_{c}^{\chi}(\alpha,x)\big\vert_{x=0}=0$. The complex
conjugate of Eq.~(\ref{eq:c-eq}) requires that $H_{c}^{\chi}(-\alpha,x)=H_{c}^{\chi}(\alpha,x)^{*}$.
Assuming the initial state is the eigenstate of $\sigma_{x}$ with
eigenvalue $\chi=+1$, i.e., $c_{+}(0)=1$ and $c_{-}(0)=0$,
we have $c_{+}(x)=H_{c}^{+}(\alpha,x)$ and $c_{-}(x)=-i\sin xH_{c}^{-}(-\alpha,x)$.
The wave function normalization further requires that $|H_{c}^{+}(\alpha,x)|^{2}+\sin^{2}(x)|H_{c}^{-}(\alpha,x)^{2}=1$.

\begin{figure}[htbp]
\includegraphics[width=8cm]{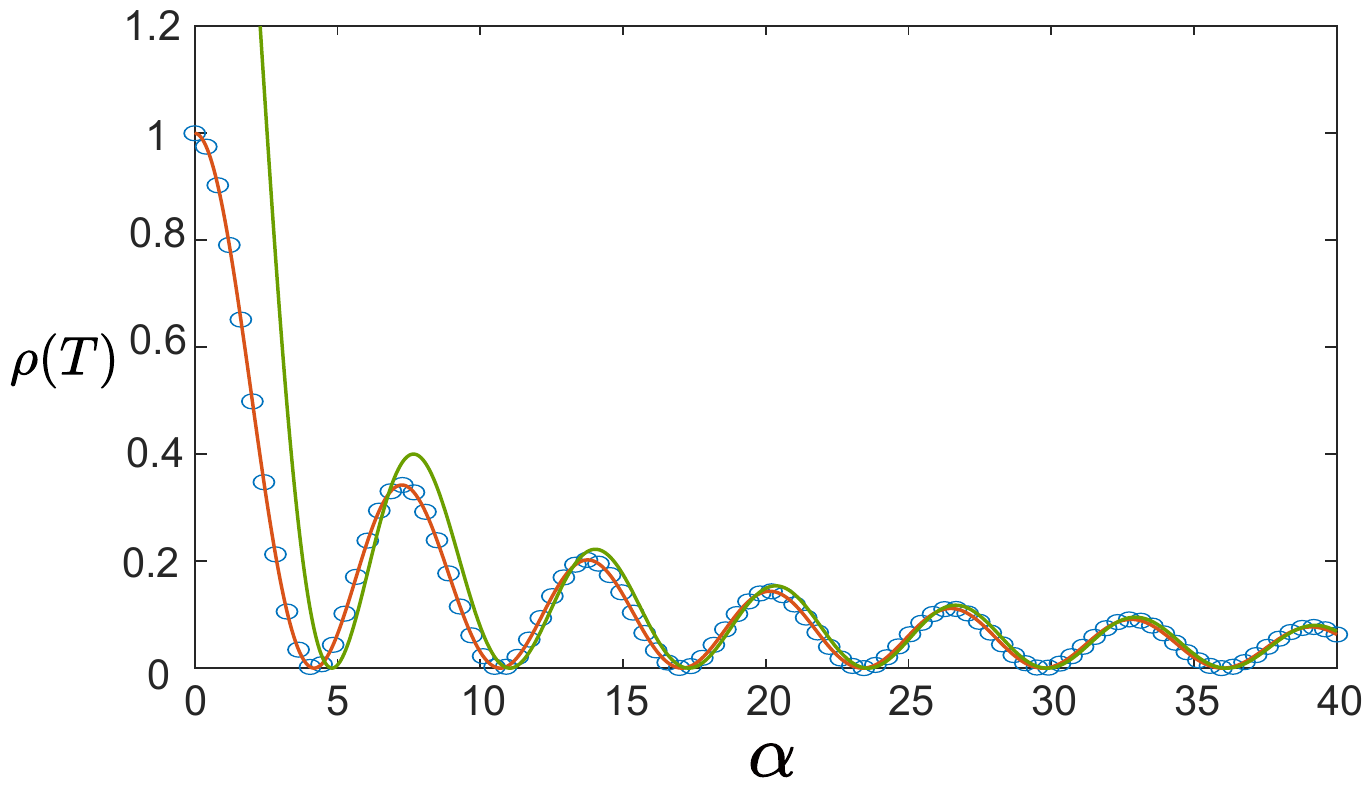} \caption{\label{Fig:alpha} The remaining probability $\rho(T)$ after one
period of evolution with different value of $\alpha$. The red line plots $H_{c}^{-}(\alpha,\frac{\pi}{2})$, while the green line plots  $\frac{\pi}{2}J_{0}(\frac{\alpha}{2})$ which is an asymptotic approximation of the former. Finally, the blue dots are the value of numerical methods.
}
\end{figure}

The solution in Eq.~(\ref{eq:exact-c-solution}) reveals a subharmonic
behavior of dynamical evolution to the $H(t)$ at a series of the
specific ratio $\alpha$, i.e., 
\begin{equation}
\Psi(t+2T)=-\Psi(t),\label{eq:2T-period}
\end{equation}
which is a hallmark feature of a DTC. Although the instantaneous eigenstates
are M\"{o}bius \ twisted and only return to themselves after $2T$, the
quench dynamics starting from a generic initial state usually does
not follow the same behavior. To analyze this question, we introduce
the time evolution operator $U(t)$, which satisfies $\Psi(t)=U(t)\Psi(0)$.
By comparing the initial condition $c_{\chi}(0)=\phi_{\chi}^{\dagger}(0)\Psi(0)$
with the exact solution in Eq.~(\ref{eq:exact-c-solution}), we can
obtain the general expression for the time evolution operator $U(t)$~\citep{sup}.
Consider the initial state $\phi_{+}(0)$. After one period $t=T$
or $x\equiv\frac{\omega T}{4}=\frac{\pi}{2}$, the probability that
the system still stays in the initial state is given by 
\[
\rho(T)=|\phi_{+}^{\dagger}(0)U(T)\phi_{+}(0)|^{2}=|H_{c}^{-}(\alpha_{n},\frac{\pi}{2})|^{2}. 
\]
 Specifically, $H_{c}^{-}(\alpha,\frac{\pi}{2})$ decays oscillatorilly
with $\alpha$, and vanishes for several $\alpha=\alpha_{n}$,
$H_{c}^{-}(\alpha_{n},\frac{\pi}{2})=0$ as shown in Fig.~\ref{Fig:alpha}.
According to the normalization condition, the other Heun function
satisfies $\left|H_{c}^{+}(\alpha_{n},\frac{\pi}{2})\right|=1$. Hence,
we can write $H_{c}^{+}(\alpha_{n},\frac{\pi}{2})\equiv\exp[i\theta_{n}]$.
The first ten $\alpha_{n}$ and $\theta_{n}$ are listed in Table~\ref{Table}.
For a large $n$, $\alpha_{n}\simeq2n\pi-\frac{\pi}{2}$ and $\theta_{n}\simeq\frac{\pi}{\alpha_{n}}$.
Consequently, we have $c_{\pm}(\frac{\pi}{2})=\exp[\pm i\theta_{n}]c_{\pm}(0)$,
indicating that the coefficients gain extra phases in one period of
evolution. This result shows that the state does not return to the
initial one after one period. In this case, the time evolution operator has the form
of 
\begin{align}
U_{n}(T) & =-i\sigma_{z}e^{-i\sigma_{x}(\frac{\alpha_{n}}{2}-\theta_{n})}
\end{align}
for specific values of $\alpha=\alpha_{n}$. For $t=2T$, it follows
that $U_{n}(2T)=-1$. Thus, after a two-period evolution, the quantum
state will return to its initial state, and gains a $\pi$ Berry phase
as shown in Eq.~(\ref{eq:2T-period}). This manifests the presence
of the DTC.

\begin{table}[htbp]
\begin{tabular}{|c|c|c|c|c|c|c|c|}
\hline 
$n$ & 1 & 2 & 3 & 4 & 5 & 6 & 7\tabularnewline
\hline 
\hline 
$\alpha_{n}$ & 4.21 & 10.73 & 17.11 & 23.44 & 29.75 & 36.07 & 42.36\tabularnewline
\hline 
$\theta_{n}$ & 0.391 & 0.199 & 0.138 & 0.108 & 0.0889 & 0.0756 & 0.0664\tabularnewline
\hline 
$n$ & 8 & 9 & 10 & 11 & 12 & 13 & 14\tabularnewline
\hline 
$\alpha_{n}$ & 48.66 & 54.95 & 61.24 & 67.46 & 73.76 & 80.07 & 86.34\tabularnewline
\hline 
$\theta_{n}$ & 0.0592 & 0.0535 & 0.0483 & 0.0459 & 0.0424 & 0.0396 & 0.0371\tabularnewline
\hline 
\end{tabular}\caption{\label{Table} The numerical value of first ten $\alpha_{n}$ of the
Heun function and the phase factor $\theta_{n}.$ With $n\to\infty$,
they have asymptotic behavior as $\alpha_{n}\simeq2\pi n-\frac{\pi}{2}$
and $\theta_{n}\simeq\frac{\pi}{\alpha_{n}}.$}
\end{table}

When the period is long ($\alpha\gg1$), we can approximate the confluent
Huen function by the zeroth Bessel function, $H_{c}^{-}(\alpha,\frac{\pi}{2})\simeq e^{i\frac{\alpha}{2}}\frac{\pi}{2}J_{0}(\frac{\alpha}{2})$,
as shown in Fig.~\ref{Fig:alpha}. The time evolution operator then
has the form, 
\begin{equation}
U(T)\simeq \left[-i\sigma_{z}\sqrt{1-\frac{\pi^{2}}{4}J_{0}^{2}(\frac{\alpha}{2})}+e^{i\frac{\alpha}{2}}\frac{\pi}{2}J_{0}(\frac{\alpha}{2})\right]e^{-i\sigma_{x}(\frac{\alpha}{2}-\frac{\pi}{\alpha})}.
\end{equation}
 If the initial state is one of the eigenstates of $\sigma_{x}$,
say $\phi_{+}(0)$, the state will mainly evolve into the other eigenstate
$\phi_{-}(0)$, and the probability to stay in the state $\phi_{+}(0)$
is just $\rho(T)$. Interestingly, the energy crossing here invalidates
the adiabatic theorem. Furthermore, after a two-period evolution,
we have $U(2T)\simeq-1$. In this limit, the system is still approximately
a DTC even if $\alpha$ is not equal to $\alpha_{n}$.

\begin{figure}{H}
\includegraphics[width=8.5cm]{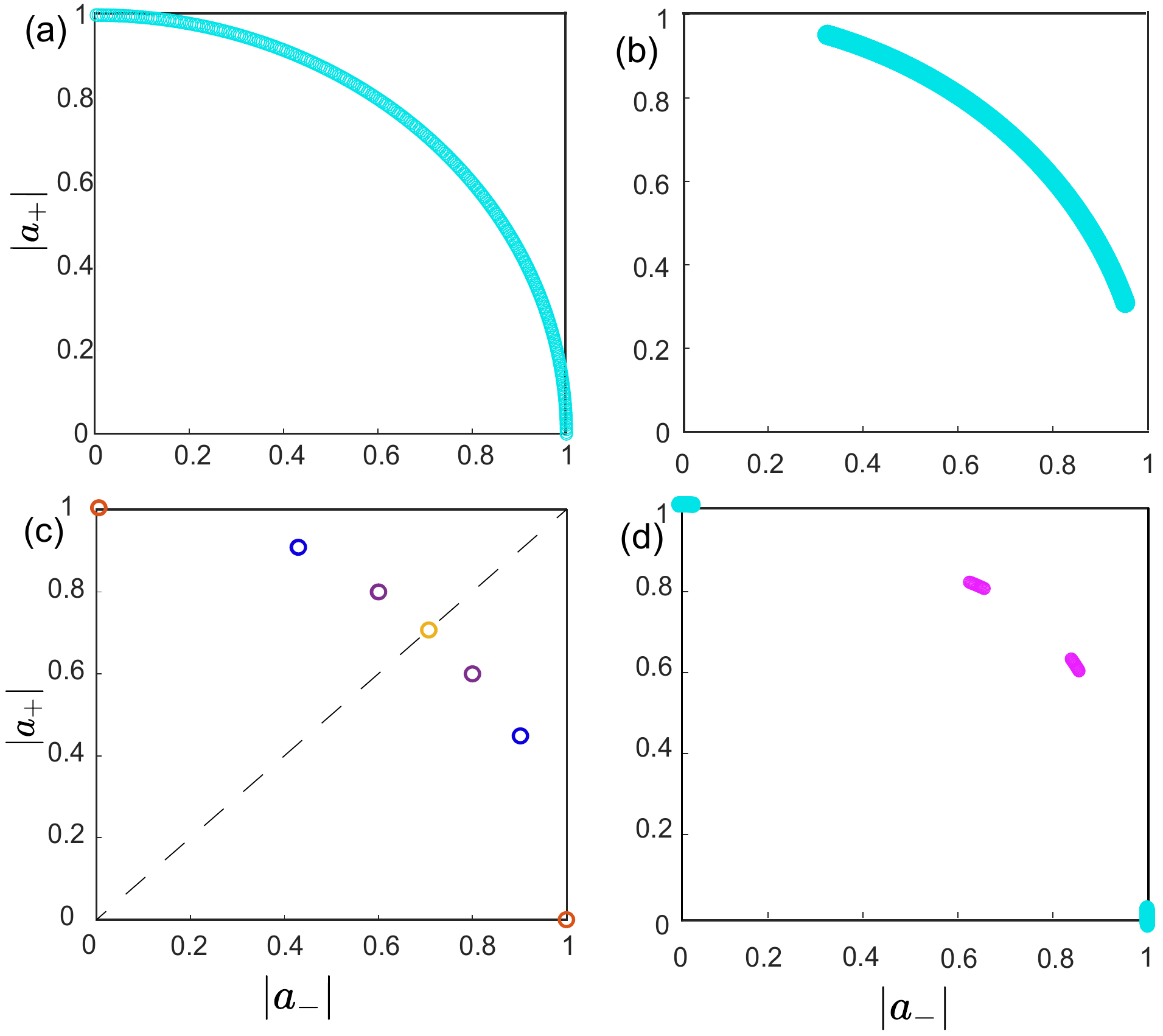} \caption{\label{Fig:Projection} The stroboscopic projection values in the
first 300 periods. (a) The projections values at $\alpha=8$ (not
one of $\alpha_{n}$) for the initial state $(1,0)^{\mathsf{T}}$,
indicating no sign of periodicity in the time evolution. (b) The projection
values at $\alpha=8$ for $(0.6,0.8)^{\mathsf{T}}$. (c) The projection
values at $\alpha=\alpha_{3}=17.11$. The different color paired points
indicate for the different initial states, $(1,0)^{\mathsf{T}}$ (red),$(0.6,0.8)^{\mathsf{T}}$
(purple), $(0.90,0.436)^{T}$(blue), and $(\sqrt{2}/2,\sqrt{2}/2)^{\mathsf{T}}$.
(d) The projection values at $\alpha=80$ (large, but not one of $\alpha_{n}$).
The initial state: $(1,0)^{\mathsf{T}}$(blue), and $[0.6,0.8]^{\mathsf{T}}$(pink).}
\end{figure}

To illustrate the periodicity of the evolution of the quantum state,
we project the state $\Psi(t=nT)\equiv U(nT)\Psi(0)$ onto the two
basis states $\phi_{\chi}(0)$, where $n$ is a positive integer.
We denote the two resulting projection parameters as $a_{\pm}(n)=|\langle\phi_{\pm}(0)\vert\Psi(nT)\rangle|$,
which satisfies $a_{+}^{2}(n)+a_{-}^{2}(n)=1$ because of the wave
function normalization. Consequently, the points $(a_{+}(n),a_{-}(n))$
should all locate on a quarter circle with a unit radius. The results
for different values of $\alpha$ and the initial states are plotted
in Fig.~\ref{Fig:Projection}. For general $\alpha$ and initial
states, the time evolution of the state does not exhibit any periodicity
since all quantum states will be visited as $n$ increases. The points
$(a_{+}(n),a_{-}(n))$ are expected to fill the entire quarter circle,
indicating ergodicity for an arbitrary initial state as shown in Fig.~\ref{Fig:Projection}(a).
We also find that there are limitations to the ergodicity because
a portion of states near the axis will not be visited for some specific
initial states and the value of $\alpha$ after a large number $n$
periods of evolution, as shown in Fig.~\ref{Fig:Projection}(b).
However, for $\alpha=\alpha_{n}$, we note that $\Psi(t=nT)$ just
have two states: $\Psi(t=nT)=\pm\Psi(0)$ for an even $n$, and $\Psi(t=nT)=\pm U(T)\Psi(0)$
for an odd $n$. The state $\Psi(t=nT)$ just bounces back and forth
between $\Psi(0)$ and $U(T)\Psi(0)$, as shown in Fig.~\ref{Fig:Projection}(c).
The period of the state becomes $2T$, indicating the formation of
a DTC. If the initial state happens to be one of the eigenstates of
$U(T)$, we have $U(T)\Psi(0)\varpropto\Psi(0)$. The evolution period
is just the same as that of the Hamiltonian. In this case, the time
translation symmetry is respected. For $\alpha$ slightly deviating
from the $\alpha_{n}$ or $\alpha\gg1$, as shown in Fig.~\ref{Fig:Projection}(d),
the system approximately keeps the two-period oscillation for a relatively
long time.

\paragraph*{Stability of the DTC}

Upon deviating from the $\alpha_{n}$, the quantum states no longer
return to their initial state after two cycles of evolution, causing
the off-diagonal term of $\sigma_{x}$ to dominate and inevitably
leading to the dynamical chaos of relevant observables. As illustrated
in Fig.~4(a), we depict $\langle\sigma_{x}(t)\rangle=\langle\Psi(t)\vert\sigma_{x}\vert\Psi(t)\rangle$
for the noninteracting model when $\alpha$ diverges from $\alpha_{n}.$
Notably, the off-diagonal term of $\sigma_{x}$ in the instantaneous
eigenstate basis can be represented as 
\begin{equation}
\langle\phi_{+1}\vert\sigma_{x}\vert\phi_{-1}(t)\rangle=e^{-i\alpha\sin^{2}\frac{\omega t}{4}}\sin\frac{\omega t}{2}.
\end{equation}
With a large $\alpha\gg1$, the dynamical phase factor $e^{-i\alpha\sin^{2}\frac{\omega t}{4}}$
prompts rapid oscillation in the observable $\text{\ensuremath{\langle\sigma_{x}(t)\rangle}}$.
Furthermore, multiple peaks emerge in the Fourier spectrum of $\langle\sigma_{x}(t)\rangle$.
Thus no time translation symmetry is respected in this case.

We demonstrate that incorporating many-body interactions can stabilize
the subharmonic response, thereby achieving a prethermal DTC. This
prethermal DTC is expected to endure in the thermodynamic limit and
exhibit robustness against perturbations and imperfect single-spin
driving fields. Specifically, we consider a spin chain by comprising
multiple identical Floquet two-level spin system with uniform nearest-neighbor
Ising couplings. The Hamiltonian can be expressed as 
\begin{equation}
H(t)=H_{0}(t)+J\sum_{i}\sigma_{\mu}^{i}\sigma_{\mu}^{i+1},\label{Eq:Many-bodyModel}
\end{equation}
with $H_{0}(t)$ given by Eq. \eqref{eq:Ham}.
Here $J$  represents
the Ising interaction between nearest-neighbor sites, $\sigma_{\mu}^{i}$
denotes the spin operator of the $i$-th site, and $\mu$ takes the values of $x,y,$or $z$. 
We assume that the interaction strength
$J$ is considerably smaller than the driving term and system's frequencies, such that $|J|\ll\hbar\omega$.
This interaction induces decoherence and thermalization of the Floquet
system~\citep{Haldarprb18,AbaninRMP19}. 
If the system is initialized 
in one of the instantaneous eigenstates of $H_{0}(t)$, decoherence
among subsystems suppresses the rapid oscillation caused by off-diagonal
terms, resulting in a smoother oscillation pattern. To demonstrate
the interaction's impact, we assess the average value of $\sigma_{x}^{i}$
for all sites, i.e., $\langle m_{x}(t)\rangle=\frac{1}{N}\sum_{i}\langle\sigma_{x}^{i}(t)\rangle,$
and display the oscillation of $\langle m_{x}(t)\rangle$ over the
initial $40$ periods. Considering Ising interactionsin the $\mu=x$ direction, 
we find that the model exhibits a persistent oscillation with a $2T$
period, even when $\alpha$ deviates from $\alpha_{n}$. 
The DTC reinforced by nonsymmorphic symmetry is resilient
to various initial states and site-resolved measurements \citep{sup}.
The oscillation's Fourier spectrum also presents an isolated peak
at $\frac{\omega}{2}$ in Fig. 4(b). This substantiates explicitly
the existence of the DTC phase in the presence of the interaction.
To further verify that the system constitutes a prethermal DTC, we
carry out a finite-size scaling of $\langle m_{x}(t)\rangle$
at stroboscopic time $t=nT$ ($n$ is an integer), defined as $\langle m_{x}(nT) \rangle=(-1)^{n}Z(n)$,
with the initial state polarized in the x-direction and present the
results for different systems' size in Fig. 4(c). The calculated $Z(n)$
show that it drops to $0.8\sim0.9$ and maintains for a while
before starting to drop drastically to around zero, which allows us
to define the lifetime $\tau=n_{c}T$ for the DTC with different parameters. The inset of Fig. 4(c) illustrates that the lifetime
$\tau$ grows exponentially with the system size $L$ , with $\tau\sim e^{bL}$, where $b$ is a phenomenological parameter that depends on the system's parameters. 
This characteristic is a key feature of a prethermal DTC with spontaneously
broken symmetry \citep{ElsePrl16}. Remarkably, the interaction preserves
the DTC phase even when the symmetry is slightly broken by a small perturbation term. 
For comparison, we also present
the lifetime and finite-size scaling for $\alpha=\alpha_{13}$ in Fig.
4(d). It is noted that the stability of $\langle m_{x}(t)\rangle$
and the lifetime of the DTC are enhanced drastically.
Thus the interacting multiple Floquet spin system is a prethermal
DTC for $\alpha=\alpha_{n}$ and even for $\alpha$ deviating slightly
from $\alpha_{n}$ in the case of interaction.

\begin{figure}[htbp]
\includegraphics[width=8.5cm]{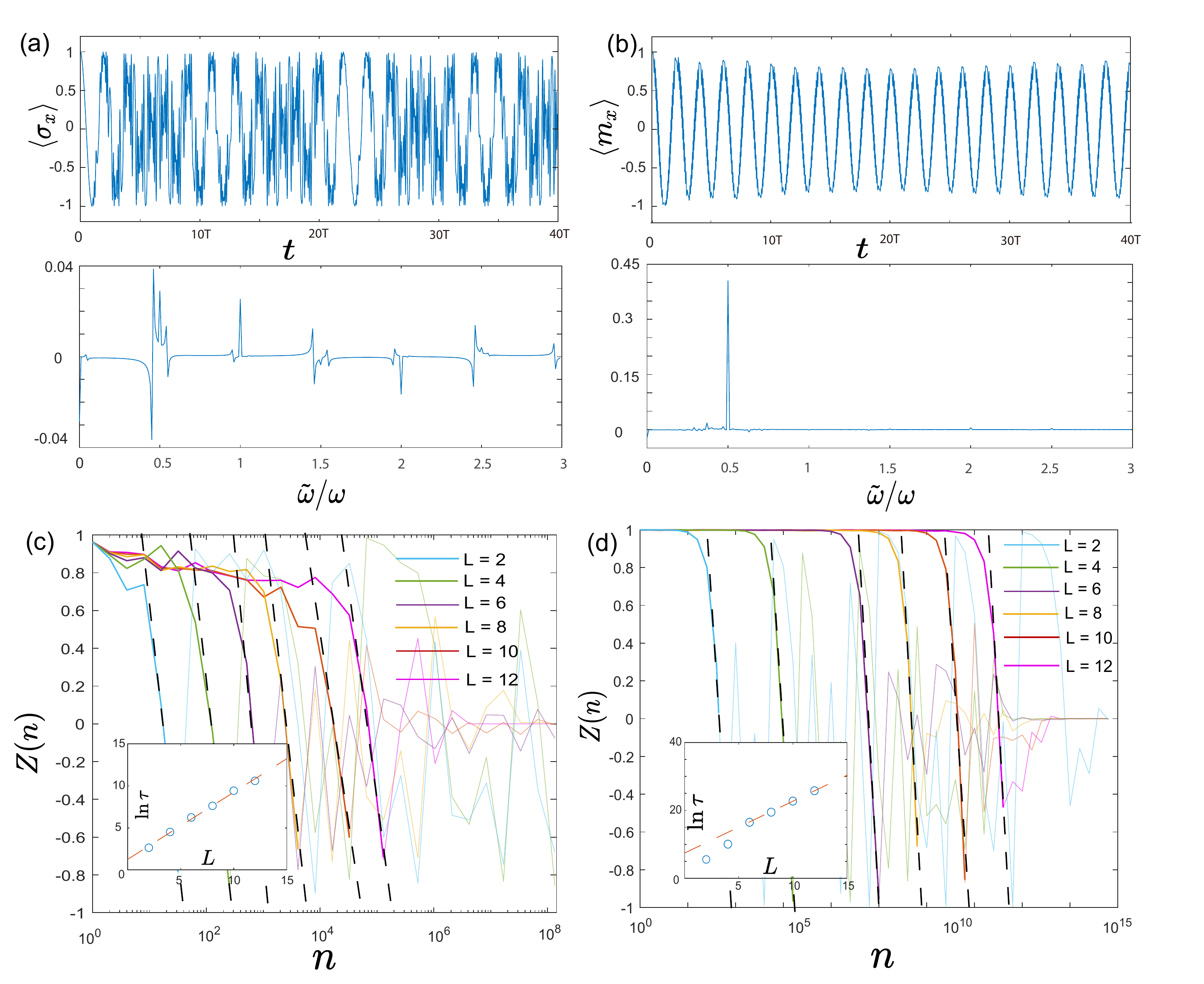} \caption{\label{Fig:Interaction} (a) $\langle\sigma_{x}(t)\rangle$ and its
Fourier spectrum for the non-interacting system for the first 40 periods.
(b) $\langle m_{x}(t)\rangle$ and its Fourier spectrum for the interacting
DTC with an initial state polarized in the
$x$-direction with length $L=10$. (c) $Z(n)$ at the stroboscopic
time in systems with varying sizes. The inset
shows the lifetime as a function of the size $L$ of the system, $\tau\propto\exp(bL),$
with $b=1.56$ for $\alpha=81.60$ (slightly deviated from $\alpha_{13}=80.07$).
(d) $Z(n)$ at the stroboscopic time for $\alpha=\alpha_{13}$ with corresponding $b=2.49$.
Here we set the interaction $J=0.2\hbar\omega$.}
\end{figure}

\paragraph*{Discussion}

In the previous section, we demonstrated that nonsymmorphic symmetry
enforces the instantaneous eigenstate into a M\"{o}bius twist, resulting in the period-doubled evolution. This period extension is attributed
to the realization of a DTC. The Hamiltonian in Eq. (\ref{eq:Ham})
possesses an additional chiral symmetry, defined as $\Gamma=\sigma_{z},$
such that $\{\Gamma,H_0\}=0$. 
The topological property of the system can be characterized by the winding number, defined as $\nu=\int\frac{dk}{2\pi}(h_{y}\partial_{t}h_{x}-h_{x}\partial_{t}h_{y})$,
if we decompose the Hamiltonian into Pauli matrices as $H_0=\bm{h}\cdot\sigma.$
When the instantaneous energy bands exhibit a single crossing, the
winding number becomes a half-integer \citep{FuNpjQ2022}. This unique
winding number is intimately connected to the system's period extension.

In contrast to gapped systems where the adiabatic theorem applies and the quantum state will always stay in instantaneous eigenstate in evolution process,
the quantum state of the DTC with band crossing returns to its initial state only after two
periods, acquiring an extra $\pi$ Berry phase when the frequency
equals to certain values or in the long-period limit.
Experimentally, quantum simulation of our DTC proposal can be performed using qubits in a quantum computer, such as quantum superconducting circuits and nuclear magnetic resonance systems~\citep{Leekscience2007,Jonesnature2000,Schroerprl2014,PhilippSciadv2022}. 
These systems enable spin manipulation for arbitrary Hamiltonians.
The period extension can be directly observed by monitoring corresponding observable quantities like spin polarization. 
Additionally, the unique Berry phase in our model can also
be measured in these systems. By selecting an initial state polarized
in the $x$-direction, the $\pi$ Berry phase can be measured from
the $2T$ periodic oscillation of $\langle\sigma_{x}\rangle$ when
$\alpha=\alpha_{n}$ or in the long-period limit.

\paragraph*{Acknowledgements}

The authors would like to thank Dr. Chenjie Wang and Dr. Huanwen Wang
for useful discussions. This work was supported by the Research Grants
Council, University Grants Committee, Hong Kong under Grant Nos. C7012-21GF,
17301220, 21304720, and 11300421.

\bibliography{citations}

\end{document}